\documentclass[hyper]{JHEP3}
\usepackage{amsmath,amsfonts}
\usepackage{epsfig}
\usepackage{cite}
\usepackage{xspace}
\def\IC{\mathbb{C}}
\def\IP{\mathbb{P}}

\def\IZ{\mathbb{Z}}
\def\IR{\mathbb{R}}
\def\hk{hyperk\"ahler\xspace}

\title{Surface Operators in ${\cal N}=2$ 4d Gauge Theories}

\author{Davide Gaiotto\\
School of Natural Sciences, Institute for Advanced Study, \\
Princeton, NJ 08540, USA
\\
{\tt dgaiotto@ias.edu} }

\abstract{${\cal N}=2$ four dimensional gauge theories admit interesting half BPS  surface operators preserving a $(2,2)$ two dimensional SUSY algebra. Typical examples are $(2,2)$ 2d sigma models
with a flavor symmetry which is coupled to the 4d gauge fields. Interesting features of such 2d sigma models,  such as (twisted) chiral rings, and the $tt^*$ geometry, can be carried over to the surface operators, and are
affected in surprising ways by the coupling to 4d degrees of freedom. We describe in detail a relation between the parameter space of twisted couplings of the surface operator and the Seiberg-Witten geometry
of the bulk theory. We discuss a similar result about the $tt^*$ geometry of the surface operator. We
predict the existence and general features of a wall-crossing formula for BPS particles bound to the surface operator.}

\begin{document}

\bibliographystyle{utphys}
\section{Introduction and outline}
${\cal N}=2$ gauge theories in four dimensions are a rich theoretical playground.
A large class of them can be conveniently engineered as the compactification of a $(2,0)$ six-dimensional SCFT on a punctured Riemann surface C.
This is a powerful tool to describe and compute protected quantities. Examples include
the gauge coupling parameter space and S-duality group, the massless effective Lagrangian \cite{Witten:1997sc} \cite{Gaiotto:2009we}, the spectrum of  BPS particles and the effective Lagrangian for the
theory on $\IR^3 \times S^1$ \cite{Gaiotto:2009hg}, the $S^4$ and the instanton partition functions \cite{Alday:2009aq}, the S-duality action on BPS line operators \cite{Drukker:2009tz}, the expectation value of
BPS line operators on $S^4$ \cite{Drukker:2009id} \cite{Alday:2009fs}.

The range of applicability of the 6d engineering approach is not fully explored. One may wonder if
it could provide a sort of classification of ${\cal N}=2$ gauge theories in four dimensions.
There are two obvious obstructions: the 6d construction might not be surjective, and is surely not injective
in the space of 4d theories.
Some known gauge theories, notably the superconformal quivers in the shape of an
exceptional Dynkin diagram solved in \cite{Katz:1997eq}, have no known 6d engineering construction.
It is possible that such a construction might yet be found, maybe involving
a Riemann surface C with the orbifold points. This was the case for
superconformal quivers in the shape of a D-type Dynkin diagram \cite{Kapustin:1998fa}.
Furthermore, the same four dimensional theory often admits several distinct six-dimensional
realizations.

It would be interesting to understand how to define an ``inverse map'', an algorithm to
identify a six dimensional ancestor for a given 4d theory. It would be even better to be able derive
the above mentioned results directly in four dimensions,
without invoking an higher dimensional construction. We believe that the
the recent work \cite{Alday:2009fs} offers a crucial clue, in the form of a certain ``minimal'' half BPS surface operator. The minimal surface operators descend from
a natural surface operator in the $(2,0)$ six-dimensional SCFT.
The 6d surface operator sits at a point in the internal Riemann surface C. As
a consequence, C coincides with the parameter space of the minimal surface operator.
To be precise, the surface operator preserves $(2,2)$ SUSY in two dimensions,
and C coincides with the parameter space of couplings in the 2d twisted superpotential.
The minimal surface operators have a set of massive vacua which, fibered over the parameter space C, produce a second curve $\Sigma$ which coincides with the Seiberg-Witten curve of the 4d theory.

These two facts are rather natural from a six dimensional point of view,
but become rather striking as soon as one identifies the minimal surface operator
as a specific defect in the 4d gauge theory, and forgets about the 6d engineering construction.
Different 6d realizations of the same 4d theory correspond to different
choices of defects in the 4d theory.

In section \ref{sec:curve} we aim to show that similar facts are universally true
for any ${\cal N}=2$ 4d theories and any possible choice of surface operators:
the twisted parameter space of the surface operator and the space of discrete
vacua fibered over it encode the Seiberg-Witten geometry of the bulk theory.

One interesting property of massive $(2,2)$ two dimensional theories are the $tt^*$ equations
(topological-antitopological fusion). The extension to surface operators in 4d theories turns out to be quite interesting, and is detailed in section \ref{sec:tt}. We mentioned that the six dimensional engineering of 4d theories was used in \cite{Gaiotto:2009hg} as a tool to understand the spectrum of massive BPS particles.
A crucial role was played by a system of Hitchin equations on C, whose spectral curve coincides with
the Seiberg-Witten curve. We will show how such Hitchin systems arise
generically in any 4d theory, given a choice of a non-trivial
surface operator, from the 2d $tt^*$ equations.

We observe that these equations control both the 4d BPS spectrum in the bulk and
the 2d BPS spectrum of particles bound to the surface
operator, generalizing the results of \cite{Cecotti:1992rm}.
We claim in section \ref{sec:wall} that this implies the existence of a
``2d-4d wall-crossing formula'' which combines the known 2d and 4d
formulae, and will be presented in detail in a separate publication

Section \ref{sec:examples} presents a few examples. Unfortunately,
the six dimensional engineering construction is the only systematic method we
have available to solve for the properties of surface operators, and covers most natural choices of
4d theories and surface operators. This makes it rather hard to
find examples illustrating the full generality of our conclusions. Rather, we will illustrate
how distinct surface operators in the same theory manage to encode the same
Seiberg Witten geometry, even though their parameter space and number of vacua differ.

We conclude with some final remarks in section \ref{sec:conclusions}
\section{SUSY review}
We will work both with conformal and with asymptotically free
theories, but it is useful to start from the $SU(2,2|2)$ 4d
${\cal N}=2$ superconformal group, and identify the subgroup
preserved by a half BPS surface operator. Indeed, all the
surface operators which we will discuss are classically
conformal invariant. The bosonic subgroup of $SU(2,2|2)$ is
$SU(2,2) \times U(1)_R \times SU(2)_R$. $SU(2,2) \sim SO(4,2)$
is the 4d conformal group. A conformal invariant surface
operator wrapping $\IR^{1,1} \in \IR^{3,1}$ will preserve 2d
conformal transformations in $\IR^{1,1}$, and rotations in the
plane perpendicular to the operator. That's $SO(2,2) \times
SO(2)_s \in SO(4,2)$, which is the block-diagonal $SU(1,1)
\times SU(1,1)\times U(1)_s$ in $SU(2,2)$.  We can complete
this to a 2d superconformal subgroup $SU(1,1|1) \times
SU(1,1|1) \times U(1)_d$ embedded in the obvious block-diagonal
way in $SU(2,2|2)$. This preserves half of the bulk
supercharges, and corresponds to a defect with $(2,2)$ 2d SUSY.
\footnote{A second class of half BPS surface operators may
exist, preserving a $SU(1,1) \times SU(1,1|2) \times U(1)_d$,
i.e. $(0,4)$ SUSY in 2d. They will play no role in this paper.
One might also consider quarter-BPS surface operators,
preserving only $(0,2)$ SUSY in 2d}

The $U(1)_R$ symmetry of the four dimensional theory becomes an
R-charge in the 2d SUSY algebra, which we will conventionally
denote as the axial $U(1)_A$. The 2d vector R-charge $U(1)_V$
is a linear combination of a $U(1)$ subgroup of $SU(2)_R$ and
of the rotation generator $U(1)_s$ in the plane orthogonal to
the surface operator. $U(1)_d$ is a second linear combination
of these two. The subset of the super(conformal) charges
preserved by the line operator is the set commuting with the
action of $U(1)_d$: the charge under rotations around the
surface operator should be equal to the weight under the
$SU(2)_R$ action. Conformal symmetry, and $U(1)_R$, can be
broken by 4d Coulomb branch expectation values, mass parameters
or gauge coupling scales. The $U(1)_A$ symmetry group (and 2d
conformal symmetry) of the surface operator will be then broken
as well. Even if the bulk theory is conformal, $U(1)_A$ can
still be broken by 2d mass parameters or strong coupling
scales.

Let us denote the two ${\cal N}=1$ sets of 4d supercharges of
the ${\cal N}=2$ theory as $Q^+_\alpha$ and $Q^-_\alpha$. The
sign $\pm$ denotes the $SU(2)_R$ weight. The 4d chirality
operator is the product of the chirality operator on the plane
of the surface operator and the chirality operator orthogonal
to the plane, i.e. 2d chirality and charge under $SO(2)_s$. The
surface operator will preserve the component of the 4d chiral
spinor $Q^+_\alpha$ which has positive 2d chirality and
positive charge under $SO(2)_s$. This 2d supercharge has
positive $U(1)_V$ charge and will be denoted as $Q_L$. This
supercharge and the conjugate $\bar Q_L$ are the left-moving
supercharges in the $(2,2)$ 2d supersymmetry algebra. The
surface operator will also preserve the component of the 4d
chiral spinor $Q^-_\alpha$ which has negative 2d chirality and
negative charge under $U(1)_s$. This 2d supercharge has
negative $U(1)_V$ charge and will be denoted as $Q_R$. This
supercharge and the conjugate $\bar Q_R$ are the right-moving
supercharges in the $(2,2)$ 2d supersymmetry algebra.

There are two  related ways to construct supersymmetric
defects in gauge theory. A simple approach is to add to the
Lagrangian terms which are integrated on the defect only. They
will include kinetic and potential terms for the degrees of
freedom on the defect, and couplings to the bulk degrees of
freedom. If the bulk Lagrangian has a superspace formulation,
the defect will break translations along half of the superspace
directions as well, and the defect Lagrangian can be written as
an integral over the unbroken superspace directions. Bulk
superfields will decompose into a tower of superfields for the
restricted defect superspace.

There is an elegant
perspective which simplifies the derivation of defect
Lagrangian, and furthermore allows one to incorporate naturally
a breaking of the bulk gauge group at the defect. To describe a dimension $d$ defect, one simply rewrites
the bulk theory as a theory in $d$ dimensions, whose fields are
valued in the space of functions of the coordinates transverse
to the defect, and whose gauge groups are maps from the
transverse space to the original gauge groups. The bulk +
defect Lagrangian is just the most general supersymmetric $d$
dimensional Lagrangian coupling this peculiar version of the bulk fields with the defect
degrees of freedom.

A crucial role in this paper is played by protected terms in
the 2d Lagrangian, i.e. superpotential or twisted
superpotential terms, and by their dependence on the bulk
fields. It turns out that bulk vector multiplets enter twisted
superpotential terms, while bulk hypers enter superpotential
terms. Indeed the scalar component of a 4d vector multiplet is
annihilated by both sets of antichiral supercharges. As they
have opposite $U(1)_V$ charge, the restriction of the scalar to
the surface operator is the lowest component of a twisted
chiral multiplet. In all examples we will consider, the 2d
degrees of freedom are massive in the IR, and the the protected
couplings of vector multiplets to a surface operator are
encoded in a 2d twisted effective superpotential ${\cal
W}(a,z)$ (See \cite{Ooguri:1999bv} for a beautiful supergravity
example). The twisted superpotential depends on the
(twisted)couplings $z_a$ of the surface operator and on the
Coulomb branch parameters of the 4d theory. It plays a role
akin to the effective prepotential for the bulk 4d theory.

The 4d hypermultiplet scalars sit in a doublet of $SU(2)_R$.
Consider the complex scalar with positive $SU(2)_R$ weight
$q^+$. It is annihilated by $Q^+_\alpha$ and by the conjugate
of $Q^-_\alpha$. Restricted to the surface operator it plays
the role of a chiral multiplet, annihilated by the supercharges
with positive $U(1)_V$. Hence it can enter the 2d
superpotential. These couplings are mostly relevant for the behavior of surface
operators in the Higgs branch of the bulk theory, though they can play a role in
the Coulomb branch of the bulk theory as well \cite{Alday:2009fs}

\subsection{4d gauge theory in 2d language}
In order to study the surface operators in
a 4d gauge theory, one simply recasts the 4d theory in a 2d language,
as a 2d gauge theory whose gauge group is the group ${\cal G}$
of maps from the transverse plane (parameterized by $x^2,x^3$)
to the original 4d gauge group $G$. The trace for the gauge
group ${\cal G}$ includes the integral over the $x^2,x^3$
directions. The $0,1$ components of the 4d connection take the
form of a ${\cal G}$ valued connection, sitting in a $(2,2)$
vector multiplet together with half of the 4d gauginos and the
4d complex scalar. The 4d gauge field in the $2,3$ directions,
or better the covariant derivative $D_2 + i D_3$ sits in a
$(2,2)$ chiral multiplet transforming under ${\cal G}$. The
moment map for the ${\cal G}$ action coincides with the
transverse field strength $F_{23}$. 4d hypermultiplets also
give rise to pairs of 2d chiral multiplets in conjugate
representations of ${\cal G}$.

The 4d gauge coupling $\tau$ plays a double role: 2d gauge
coupling and Kahler parameter for the $D_2 + i D_3$ chiral
field. The various pieces of the 4d gauge kinetic term arise
from the 2d kinetic terms. In particular, the $F_{23}^2$ term
arises from integrating away the 2d auxiliary field D, which
couples to the moment map $F_{23}$. The 4d kinetic energy for
4d hypermultiplets is a combination of the 2d kinetic energy, a
superpotential term involving the $D_2 + i D_3$ derivative of
the hypermultiplets, and the D term potential.

The advantage of this construction is that it makes rather
simple to add a surface operator along the $0,1$ directions.
For example we can define a surface operator by adding extra 2d
chiral multiplets, and possibly 2d gauge fields, to the setup.
The overall moment map for the ${\cal G}$ gauge action has an
extra contribution from the moment map of the 2d matter
$\mu^{2d}$, and takes the form $F_{23} + \delta(x^2)
\delta(x^3) \mu^{2d}$. We see how SUSY, or more precisely the
D-term constraints, force the connection to have a monodromy
around the surface operator (see \cite{Gukov:2006jk} for the
corresponding statement in ${\cal N}=4$ SYM). Notice also that
the 2d F-term and D-term equations in presence of
hypermultiplets coincide with the equations for BPS vortices
localized in the $2,3$ directions. This is no coincidence.
Surface operators and normalizable vortex solutions are clearly
related. A vortex in a Higgs branch or a mixed Higgs-Coulomb
branch of a ${\cal N}=2$ theory will flow in the IR to a
surface operator in the IR theory. Many beautiful results about
vortex operators (see for example the \cite{Tong:2008qd}
review) can be recast in the language of surface operators.

In two dimensions the effective twisted superpotential receives
contributions from 2d instanton (``vortex'' in a different sense) configurations,
where the chiral fields are covariantly holomorphic, and the
magnetic flux is set to be equal to the moment map. In the 4d
setup, 4d instantons are a neat example of a 2d instanton for
the ${\cal G}$ 2d gauge theory (see also \cite{Atiya}): $D_2 + i
D_3$ should be covariantly holomorphic and $i F_{01} = F_{34}$.
The 4d gauge coupling enters the instanton action as a 2d
Kahler parameter for the $D_2 + i D_3$ chiral field. Coupling
to 2d matter allows one to study combinations of 2d vortices
and 4d instantons, where the moment map for the 2d matter acts
as a source for the $i F_{01} = F_{34}$ selfduality condition.
We will see from explicit example that these mixed 4d-2d
instantons indeed appear correct the effective twisted
superpotential and twisted chiral ring relations.

\section{Surface operators in the Abelian IR theory}
\subsection{Seiberg-Witten geometry from surface operators} \label{sec:curve}
We will make the following assumptions about the IR behavior of
the theory: \begin{itemize} \item The massless degrees of
freedom of the 4d theory consist of an abelian gauge theory of
rank $r$ \\ \item There are no 2d massless degrees of freedom:
the 2d theory is massive. (Notice that the Coulomb branch
parameters of the 4d theory enter the 2d Lagrangian as twisted
masses, so this assumption is not very restrictive.)
\\ \item The 2d surface operator has a finite number of vacua,
parameterized by the expectation values of the operators in the
2d twisted chiral ring, subject to the twisted chiral ring
relations. The structure constants of the ring may depend on
the parameters $z_a$ and 2d twisted masses, on the 4d Coulomb
branch parameters, gauge couplings and mass parameters. \\ \item The
parameters $z_a$ of the surface operator can be varied by
adding a term $\delta z_a \hat x_a$ to the twisted superpotential.
$\hat x_a$ are appropriate elements of the twisted chiral ring. There might be a space of
marginal superpotential deformations as well, but it will play no role in the following. We will denote the space of 2d parameters as ${\cal P}$.
\end{itemize}
As long as the surface operator has a good UV definition,
 ${\cal P}$ can depend on the UV gauge couplings of the 4d
theory, but not on the 4d Coulomb branch parameters or masses. The
2d IR vacua of the surface operator can be fibered over ${\cal
P}$ to give a new manifold $\hat {\cal P}$. $\hat {\cal P}$
will in general depend on the 4d Coulomb branch parameters and
masses. The expectation values $x_a$ of the $\hat x_a$
operators will give a useful one form $\lambda = x_a dz_a$ on
$\hat {\cal P}$. The six-dimensional construction gives a
canonical example of this setup: ${\cal P}$ coincides with the
curve $C$, $\hat {\cal P}$ with the Seiberg-Witten curve and
$\lambda$ with the Seiberg-Witten differential.

As the 2d theory is, by assumption, massive, in the IR the
surface operator takes the form of a defect in the 4d IR
abelian gauge theory. The defect is characterized by the
parameters $\alpha^i$ and $\eta_i$, where $\alpha^i$ are the
monodromies of the abelian gauge fields around the operator,
and $\eta_i$ the 2d theta angles, couplings to the magnetic
fluxes on the surface. The two sets of parameters are both
angular variables, and are exchanged by the abelian
electric-magnetic duality group \cite{Gukov:2006jk}. The angles are
naturally combined in coordinates $t_i = \eta_i + \tau_{ij}
\alpha^j$ on an Abelian variety. This Abelian variety with
complex structure given by the IR bulk gauge couplings is a
familiar object in Seiberg-Witten theory.

What is the relation between the IR parameters $t_i$ and the
original parameters $z_a$ of the surface operator? The IR
couplings are encoded in an IR effective twisted
superpotential, as
\begin{equation}
t_i =\frac{\partial {\cal W}_{eff}}{\partial a^i}
\end{equation}
Indeed the superspace integral of ${\cal W}_{eff}[a^i]$ gives
the 2d Lagrangian coupling
\begin{equation}
\frac{\partial {\cal W}_{eff}}{\partial a^i} F_+^i + c.c
\end{equation}
Here $F_+^i$ is the restriction to the surface operator of the
self-dual field strength.

The effective twisted superpotential should also control the
expectation value $x_a$.
\begin{equation}
x_a =\frac{\partial {\cal W}_{eff}}{\partial z_a}
\end{equation}
We can write
\begin{equation}
\frac{\partial t_i}{\partial z_a} =\frac{\partial x_a}{\partial a^i}
\end{equation}

Consider the periods of the differentials $\frac{\partial
x_a}{\partial a^i} dz_a=\frac{\partial \lambda}{\partial a^i}$
over a one-cycle $\gamma$ in $\hat {\cal P}$. As the cycles
starts and ends at the same point in $\hat {\cal P}$, the
variation of $t_i$ along the cycle must be a period of the Abelian variety,
of the form $n_i + \tau_{ij}m^j$ for some integers $n,m$. Alternatively, in terms
of more general coordinates $u_s$ on the 4d Coulomb branch
moduli space,
\begin{equation}
\oint_{\gamma \in \hat {\cal P}} \frac{\partial \lambda}{\partial u_s} = n_i[\gamma] \frac{\partial a^i}{\partial u_s}+ m^i[\gamma] \frac{\partial a^D_i}{\partial u_s}
\end{equation}

This statement strongly resembles the basic structure of the
Seiberg-Witten geometry, with $\hat {\cal P}$ playing the role
of the SW curve and $x^a dz_a$ the role of the SW differential.
In particular, this defines a map $\gamma \to (n_i,m^i)$ from
the homology of one-cycles in $\hat {\cal P}$ to the charge
lattice of the 4d theory. Up to possible integration constants,
one is tempted to guess that under this map the central charges
of the 4d theory are reproduced
\begin{equation}
Z_\gamma =\oint_\gamma \lambda
\end{equation}

Notice that the central charge function is a linear map on an
extended lattice $\hat \Gamma$, which include the gauge
charges, but also the flavor charges of the theory, which
multiply the mass parameters. These include the mass parameters
of the 4d theory, but also possibly some extra ``twisted
masses'' of the 2d theory. The above, tentative relation
would require an extended map from $H_1(\hat {\cal P},\IZ)$ to
$\hat \Gamma$.

It should be possible to derive such relation directly, and not
just in a differentiated form. One possibility is to look at
dynamical BPS domain walls on the surface operator. On general
grounds, the tension of such domain walls is given in terms of
the variation of the superpotential across the wall, i.e.
$|\Delta {\cal W}|$. If we use the expression for the effective
IR superpotential, this gives
\begin{equation}
\Delta {\cal W} =\int_{\hat z'}^{\hat z''} \lambda
\end{equation}
Here $\hat z'$, $\hat z''$ are two points in $\hat {\cal P}$ in
the fibre of the same point $z$ in ${\cal P}$. This can be
interpreted as a central charge, which receives contributions
from the topological charge of the soliton (the choice of vacua
$\hat z'$, $\hat z''$ above $z$) and possibly the gauge and
flavor charges of the soliton. More precisely, different
solitons may be associated to different choices of paths
between $\hat z'$ and $\hat z''$. The difference in their
central charge is the integral of $\lambda$ on the difference
$\gamma$ between the two paths. $\gamma$ is a closed path, and
we ascribe the difference in the central charges
\begin{equation}
Z_\gamma =\oint_\gamma \lambda
\end{equation}
to the difference in the flavor and gauge charges carried by
the two solitons.

A few observations are in order. Both the homology lattice
$H_1(\hat {\cal P},\IZ)$ and the 4d charge lattice $\hat
\Gamma$ are local systems over the Coulomb branch of the 4d
theory. $\hat \Gamma$ has monodromies around loci where
massless 4d BPS particles appear. The map from $H_1(\hat {\cal
P},\IZ)$ to $\hat \Gamma$ should intertwine the monodromies of
the two local systems. This fact almost answers an important
question regarding the image of the map $H_1(\hat {\cal
P},\IZ)$ to $\hat \Gamma$, which in principle may or not be
surjective. Indeed, as the image includes at least a non-zero
charge vector, it will have to include the linear span of all
the images of that charge under monodromy transformations. The
result is strengthened by some results we will accumulate in
the next sections: bulk 4d BPS particles can form bound states
with 2d BPS particles, much like 4d particles can form bound
states among themselves. The appearance and disappearance of
such bound states is determined by the IR gauge, flavor and
topological charges of the particles only. Each time a 4d
particle of charge $\gamma$ binds to a 2d domain wall, it
implies that $\gamma$ sits in the image of the map from
$H_1(\hat {\cal P},\IZ)$ to $\hat \Gamma$. Monodromies and
wall-crossing should be sufficient guarantee surjectivity in
non-degenerate cases.

In the case of the minimal surface operator in theories built
from six dimensions the map is indeed surjective. All the
periods of the 4d theory are reproduced as periods of $\lambda$
over the Seiberg-Witten curve. An even stronger condition is
true: the 4d Coulomb branch coincides with the space of
possible normalizable deformations of the Seiberg-Witten curve
$\hat {\cal P}$. It is unclear from the derivation in this
section if a similar statement would hold in a more general
setup. The next sections will put further constraints, but will
not provide a definitive answer.

\subsection{Hitchin systems from surface operators} \label{sec:tt}
The $tt^*$ equations are a beautiful property of $(2,2)$
theories \cite{Cecotti:1991me}, which extends well to surface
operators in ${\cal N}=2$ 4d theories. Consider the
compactification of the 4d theory on a circle of radius $R$,
with the surface operator wrapping $\IR \times S^1 \in \IR^3
\times S^1$ (and Ramond boundary conditions). Following the 2d
story, instead of the discrete fibration $\hat {\cal P}$, it is
possible to consider now a vector bundle ${\cal V}$ of Ramond
vacua over the parameter space ${\cal P}$. This bundle is
endowed with two natural structures:  an Hermitian connection
$D_a, \bar D_a$ and a holomorphic one form $c_a dz_a$ (and
$\bar c_a d \bar z_a$) valued in endomorphisms of ${\cal V}$,
given by the action of the twisted chiral operator $x_a$ over
the Ramond vacua of the theory.

The two objects satisfy a generalization of Hitchin's
equations:
\begin{align}
&[D_a,D_b]=[\bar D_a, \bar D_b]=[c_a,c_b]=[\bar c_a, \bar c_b]=0 \notag \\
&D_a c_b = D_b c_a \qquad \bar D_a \bar c_b = \bar D_b \bar c_a \qquad D_a \bar c_b = 0 \qquad  \bar D_a c_b=0 \notag \\
&[D_a, \bar D_b] + [c_a, \bar c_b]=0
\end{align}

These are equivalent to the flatness of the spectral connection
\begin{equation}
\nabla_a = D_a + \frac{R}{\zeta} c_a \qquad \bar \nabla_a = \bar D_a + R\zeta \bar c_a
\end{equation}
for all values of $\zeta$. These results admit a very simple
and intuitive derivation in terms of supersymmetric Janus
configurations, which is included in appendix \ref{app:tt}.

The relation between ${\cal V}$ and $\hat {\cal P}$ is
straightforward: the $c$ endomorphisms which describe the
action of twisted chiral ring operators commute, and their
eigenvalues are the expectation values of the corresponding
operators on $\hat {\cal P}$. The main difference between the
pure 2d case and the case of surface operators is the fact that
$c,D$ will also depend on the 4d Coulomb branch parameters. The
effect of the 4d Coulomb branch parameters is similar to the
effect of 2d twisted masses. The effect of twisted masses on
the $tt^*$ equations has been studied before,
but the result are unpublished, and possibly lost.
\cite{cecotti}. From the very
beginning, we will see a strong resemblance with ideas
developed in the context of the 4d $tt^*$ equations
\cite{Gaiotto:2008cd}. Indeed, our final result will be a neat
merger of the 2d and 4d $tt^*$ perspectives.

Upon compactification on a circle, the Coulomb branch moduli
space of the theory doubles in dimension, as an Abelian variety
of electric and magnetic Wilson lines is fibered over the 4d
moduli space. The 3d Coulomb branch moduli space ${\cal M}$ is
an \hk manifold. It has a has a $\IC \IP^1$ worth of complex
structures. We will always label the choices of complex
structures by a inhomogeneous parameter $\zeta$. The complex
structures $\zeta=0$ or $\zeta=\infty$ are special: the 4d
Coulomb branch parameters are holomorphic in these complex
structures, and the torus of Wilson lines is an Abelian variety
(dual to the one we met in the previous section). The
holomorphic functions in other complex structures $\zeta \in
\IC^*$ are rather more interesting, and are the main subject of
the analysis of \cite{Gaiotto:2008cd}.

A canonical example of such function is the expectation value
of a half BPS line operator in the 4d theory, wrapped along the
$S^1$ \cite{line}. A BPS line operator stretched along the
$x^1$ direction preserves a linear combination $Q^\pm + \zeta
\gamma_1 \bar Q^\pm$ of the chiral and anti-chiral supercharges
(it preserves $SU(2)_R$). $\zeta$ is a pure phase for a
physical operator, but it can be analytically continued to
$\IC^*$. In terms of a low energy sigma model on the 3d moduli
space of vacua ${\cal M}$, the linear combination of
supercharges $Q^\pm + \zeta \gamma_1 \bar Q^\pm$ kills a
certain ring of protected operators, which are holomorphic
functions of the scalar fields in complex structure $\zeta$.

A free Abelian example of a BPS Maldacena-Wilson line operator
and its expectation value would have the schematic form
\begin{equation}
\langle P\exp \oint \frac{1}{2}\zeta^{-1} a + i A + \frac{1}{2} \zeta \bar a \rangle = \exp \left[\pi R \zeta^{-1} a + i \theta + \pi R \zeta \bar a\right]
\end{equation}
Here $a$ is the vector multiplet scalar
field and $\theta$ the Wilson line order parameter.
Furthermore, every mass parameter of the 4d theory (and twisted
mass parameter of the 2d theory) is associated to a flavor
symmetry, and we can include an extra flavor Wilson line
$\theta_f$ for each mass parameter $m$. It is natural to
restrict BPS operators in complex structure $\zeta$ to be
functions of the natural combination $\pi R \zeta^{-1} m + i
\theta_f + \pi R \zeta \bar m$. From now on, every time we
mention holomorphic functions on ${\cal M}$, we implicitly
assume such a dependence on the masses and flavor Wilson lines
as well. (Including both 4d masses and 2d twisted masses.
Notice that both are expectation values of a background vector
multiplet gauging the flavor symmetry.)

The spectral connection $\nabla$ will depend on the choice of
3d vacuum in ${\cal M}$ and on the mass parameters and flavor
Wilson lines. We would like to identify a sense in which the
spectral connection with given parameter $\zeta$ depends on
${\cal M}$ holomorphically in complex structure $\zeta$ (and
$R$ identified with the radius of the compactification circle).

Indeed, consider the monodromy data of the spectral connection. In
simple cases, it will consist of traces of monodromy matrices
around cycles of ${\cal P}$. In more complicated cases, it will
include Stokes data at irregular singularities. In any case,
the result is a set of functions on ${\cal M}$. We want to
argue that such functions will be holomorphic in
complex structure $\zeta \in \IC \IP^1$.

The BPS projection for a surface operator and the BPS
projection for a line operator in the 4d theory parallel to it
are compatible, and intersect on a set of two supercharges. The
same amount of supersymmetry is in general preserved by a line
operator (a non-dynamical domain wall, or a boundary) inside a
surface operator. Such line operators are analogous to
supersymmetric boundary conditions in a 2d theory, relating the
left and right moving supercharges as $Q_L + \zeta \bar Q_R=0$
and $Q_R + \zeta \bar Q_L=0$.

As proven in \cite{Hori:2000ck} and reviewed in the appendix
\ref{app:tt} the correlation function of the 2d theory on a
half-cylinder of radius $R$, with such a boundary at one end
and a choice of Ramond vacuum at the other end, is a flat
section of the spectral connection $\nabla$ with the same $R,
\zeta$! This statement can be immediately extended to
supersymmetric line operators between two different theories,
or the same theory at different values of the parameters $z$
and $z'$, by the reflection trick: an interface between two 2d
theories is the same as a boundary condition for the product
theory.

The expectation value of a line operator interpolating between
given Ramond vacua of the theory at different values $z$ on the
left and $z'$ on the right of the parameters will be a matrix
flat section $M(z,z')$. (This is a flat section for both a left
action of the spectral connection in $z$ and for a right action
of the spectral connection in $z'$.) A particularly interesting
line operator is the Janus domain wall defined in detail in
appendix \ref{app:tt}. It is defined starting from a trivial
line operator at $z=z'$ and continuously deforming the coupling
$z$ to a given final value while preserving the same SUSY.
The expectation value of the operator only depends on the
homotopy class of the path in ${\cal P}$ between $z$ and $z'$.
For the trivial line operator $M(z',z')$ is the identity
matrix, and we can use the flat spectral connection to
transport $z$ along the path as we define the Janus line
operator. Hence the expectation value of the Janus line
operator literally computes the transport matrix for the flat
connection. In particular the transport matrix is the
expectation of a line operator annihilated by the two
supercharges $Q_L + \zeta \bar Q_R=0$ and $Q_R + \zeta \bar
Q_L=0$.

The same result will apply for the case of surface operators.
The spectral data for the flat connection, which is computed
from the transport matrix, gives functions over ${\cal M}$
annihilated by the two supersymmetries preserved by the line
operator. These two supercharges are two out of the four $Q +
\zeta \gamma_1 \bar Q$ which annihilate holomorphic functions
in complex structure $\zeta$. Notice that the kernel of the two
supercharges coincides with the kernel of the full set of four
supercharges: the four supercharges in the 3d low energy sigma
model on ${\cal M}$ all have the general form $\psi^i
\partial^\zeta_i$ and only differ in the $SU(2)_R$ and spacetime indices of the
fermion $\psi$.

In the six dimensional setup in \cite{Gaiotto:2009hg}, a very
specific Hitchin system on $C$ produces holomorphic functions
in complex structure $\zeta$ as the monodromy data of the
spectral connection. We see that the $tt^*$ equations for a
generic surface operator can play a similar role. In the
six-dimensional setup, the choice of Hitchin system is
determined by the requirement that the spectral curve $\det
(x-c(z))=0$ should coincide with the Seiberg-Witten curve and
$x dz$ with the Seiberg-Witten differential. (Remember that
$c(z)$ plays the role of the Higgs field in the Hitchin
system.) This is just the expected relation between ${\cal V}$
and $\hat {\cal P}$.

In general the gauge invariant information about $\hat {\cal
P}$ encoded in the $c_a(z)$ can be packaged into a ring of
symmetric differentials of various degree $k$, $\mathrm{Tr}
(c_a dz_a)^k$ on ${\cal P}$. We are now given several, possibly
related complex manifolds: the 4d Coulomb branch, the moduli
space of possible $\hat {\cal P}$, the moduli space of
normalizable deformations of the degree $k$ differentials. They
all coincide in the six dimensional example, but we do not know
if that will be true in general. As a step in that direction,
it is useful to consider the full map from the 3d Coulomb
branch of vacua to the moduli space of solutions of the $tt^*$
Hitchin-like system. This map extends the map between the 4d
Coulomb branch and the moduli space of possible $\hat {\cal
P}$.

It is not clear to us if the moduli space of solutions of the
multidimensional $tt^*$ Hitchin-like system would admit a \hk
metric, as the equations do not have the form of a \hk quotient
in general. We do not actually know how to even define the
moduli space manifold precisely, though it should bear some relation to the
some space of Higgs bundles on ${\cal P}$. In any case, given any one-dimensional
submanifold of ${\cal P}$, we have a well-defined \hk moduli
space of solutions of Hitchin equations over it. Hence we have
a map from the 3d Coulomb branch to the moduli space of this
one dimensional Hitchin system. It is a map between \hk
manifolds which commutes with the structure of fibrations by
Abelian varieties, and is holomorphic in all complex structures
$\zeta \in \IC \IP^1$. Such maps between different \hk
manifolds are rather uncommon.

It would be interesting to explore if the moduli space of solutions of the multidimensional
$tt^*$ Hitchin system could exactly coincide in general with the 3d Coulomb
branch of the theory. For this to be true, it would have to be the case that the $tt^*$
equations for a genuinely 2d theory had no moduli space.

\subsection{Wall-crossing and surface operators} \label{sec:wall}
In two dimensions, the $tt^*$ equations are part of an
interesting structure \cite{Cecotti:1992rm}: the spectral connection
commutes with a pair of connections $\nabla_\zeta=\zeta
\partial_\zeta + {\cal A}^{2d}_\zeta$, $\nabla_R=R \partial_R +
{\cal A}^{2d}_R$. All connections have simple poles at
$\zeta=0, \infty$. In particular the $\nabla_\zeta$ connection
has irregular singularities at $\zeta=0, \infty$, which lead to
Stokes phenomena. The Stokes factors for the $\zeta$ connection
can be computed in a large radius limit (as $\nabla_R$ commutes
with $\nabla_\zeta$ and the location of Stokes rays is $R$
independent, the Stokes factors are also $R$ independent) and
turn out to be in one-to-one correspondence with the 2d BPS
particles in the theory.

The $\nabla_a$ connection commutes with $\nabla_\zeta$ as well,
but the location of the individual Stokes rays is a function of
the $z_a$ (it coincides with the phase of the central charge of
the corresponding BPS state). The product of all Stokes factors
in a wedge in the $\zeta$ plane is still invariant, as long as
no rays enters or exits the wedge. This leads to a simple wall-crossing
formula for the BPS particles of the 2d theory.

The holomorphic functions on ${\cal M}$ are governed by a
formally similar set of equations \cite{Gaiotto:2008cd}, which
we denote as 4d $tt^*$ equations. In the 4d $tt^*$ setup, one
has a compatible set of connections ${\cal A}_\zeta,{\cal
A}_R,{\cal A}_u$, along $\zeta, R$ and along the 4d Coulomb
branch moduli, for the bundle of functions of the electric and
magnetic Wilson line parameters. More concretely, ${\cal
A}_\zeta,{\cal A}_R,{\cal A}_u$ are differential operators in
the Wilson line parameters. In particular, the connection
$\partial_u + {\cal A}_u$ has the interpretation of
Cauchy-Riemann equations for a holomorphic functions on ${\cal
M}$. The Stokes data of the connection on the $\zeta$ plane
captures the BPS spectrum and wall-crossing of the 4d theory.
Rather than finite matrices, the Stokes factors take the form
of KS transformations, which are symplectomorphisms of a
certain complex torus.

In the context of surface operators we have the 2d $tt^*$
connection $\nabla_a$, and one might wonder if a connection
$\nabla_u$ along the 4d Coulomb branch might also exist on the
bundle of Ramond vacua of the surface operator, compatible with
the connection $\nabla_a$. This is cannot be the case, as the
spectral data of $\nabla_a$ depends on the 4d Coulomb branch
parameters! As the spectral data defines holomorphic functions
on ${\cal M}$, we can instead consider a combined 2d-4d
connection $\nabla_u + {\cal A}_u$. This should be seen as a
connection on the bundle of functions of the Wilson line
parameters of the 3d theory, valued in ${\cal V}$. Here and
below $\nabla$ denotes a connection valued in
endomorphisms of the finite dimensional bundle ${\cal V}$, and ${\cal
A}$ is the standard 4d connection valued in differential operators.

Similarly, we expect some $\nabla_\zeta + {\cal A}_\zeta$ and
$\nabla_R + {\cal A}_R$. The existence of such connections
is a consequence of the (possibly anomalous) scale
invariance and $U(1)_A$ symmetries of the combined 2d-4d
system. (This was true both in the 2d $tt^*$ and in the 4d
$tt^*$ separately.) It is natural to expect that the Stokes
data for the combined connection $\nabla_\zeta + {\cal
A}_\zeta$ will describe the spectrum and wall-crossing of BPS
particles bound to the surface operator and their interactions
with the 4d particles in the bulk.

It is possible for 4d BPS particles to bind to 2d BPS
solitons, giving rise to mixed 2d-4d wall-crossing formulae.
Indeed the 2d BPS particles carry 4d gauge charges, and, say, a
4d electron should be able to form bound states to a 2d
monopole. The 2d wall-crossing formula expresses the invariance
of a product of Stokes factors across the walls of marginal
stability. These 2d Stokes factors are finite matrices. The 4d
wall-crossing formula involves Stokes factors valued in a group of symplectomorphisms
of certain formal variables $x_\gamma$ labeled by the elements
$\gamma$ of the charge lattice of the 4d gauge theory. The
structure group of the 4d-2d connection appears to be a
semi-direct product of the group of symplectomorphisms of the
formal variables $x_\gamma$ and of a group of finite matrices
valued in the $x_\gamma$.

The detailed formulation and concrete examples of such 2d-4d
wall-crossing formula is left to a separate publication.

\section{Examples} \label{sec:examples}
\begin{figure}
  \begin{center}
    \includegraphics[width=5in]{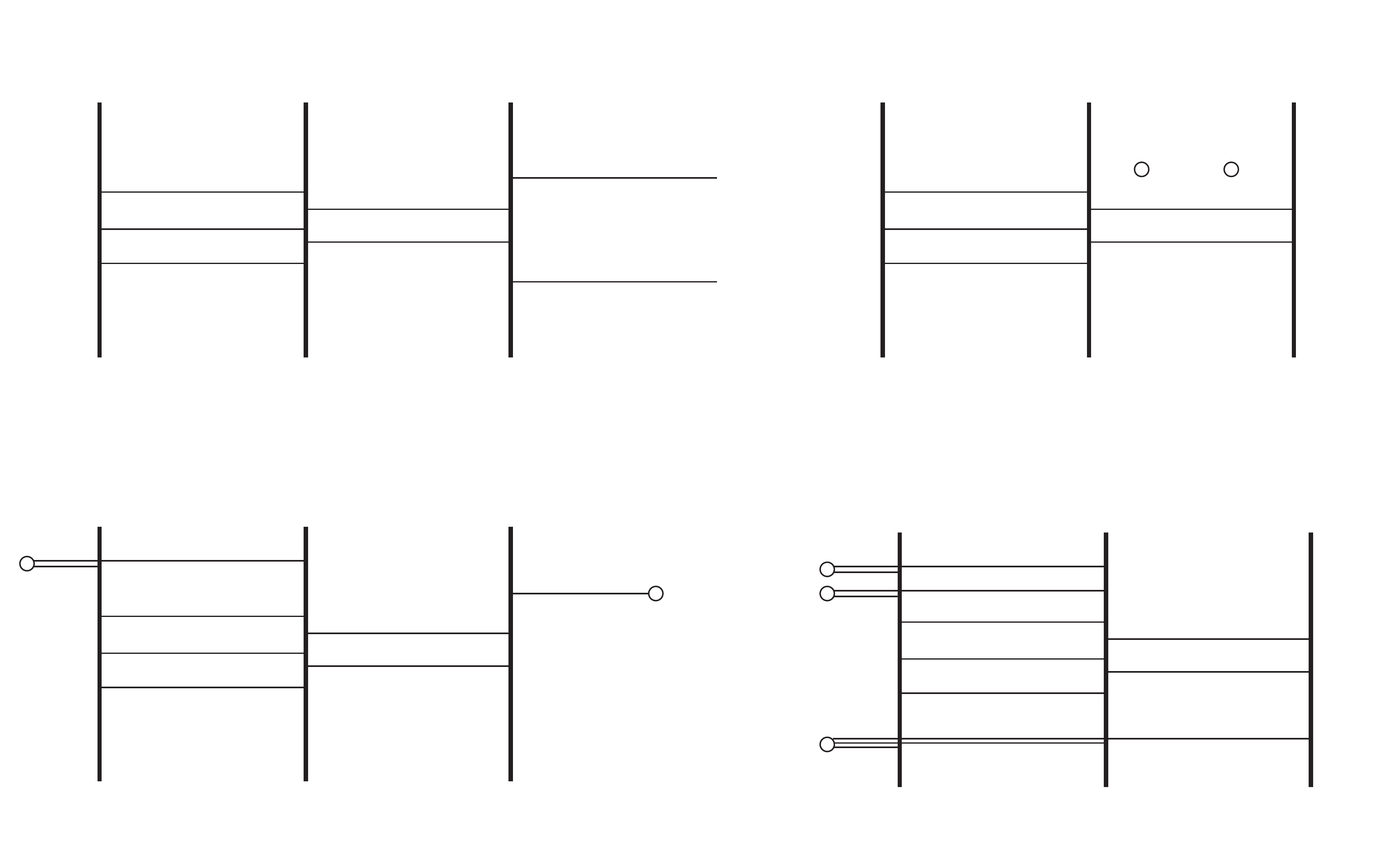}
      \end{center}
  \caption{Different brane realizations of a simple quiver gauge theory: $SU(3) \times SU(2)$ with a bifundmental and two $SU(2)$ fundamentals.
  Vertical lines represent NS5 branes, horizontal D4 branes, circles are D6 branes.
  (a) Simplest realization. Flavors from semiinfinite D4 branes. (b) Two D6 can also produce the flavors
  (c) D4 segments are created when moving the D6 branes. (d) An extra D6 has been added to the right, brought to the left.}
  \label{fig:ns5d4d6}
\end{figure}

Several of our examples are based on a simple Type IIA brane
construction introduced in \cite{Witten:1997sc} to engineer
specific ${\cal N}=2$ gauge theories, and extended by
\cite{Hanany:1997vm} to engineer two dimensional $(2,2)$ sigma
models. The construction involves an array of NS5 branes (along
the $012345$ directions) and finite or semi-infinite D4 brane
segments stretched between them (along the $01236$
direction)(See fig. \ref{fig:ns5d4d6} (a)). It may be enriched
by extra D6 branes placed in the intervals between NS5 branes
(along the $0123789$ directions)(See fig. \ref{fig:ns5d4d6}
(b)). The D6 branes may be moved from an interval to another,
without changing the gauge theory interpretation of the setup.
Hanani-Witten D4 brane creation effects play an important role
in the process (See fig. \ref{fig:ns5d4d6} (c)). Each Type IIA
construction admits infinitely many  distinct lifts to a 6d
engineering construction \cite{Gaiotto:2009hg}: the lift
requires all D6 branes to be moved to the far left or the far
right of the system, and different choices lead to different
curves C. Furthermore, any number of extra D6 branes can be
added far on one side of the system (in the absence of D6
branes or semi-infinite D4 branes on that side) and moved to
the other side (see fig. \ref{fig:ns5d4d6} (d)). In particular,
one is lead to a variety of Hitchin systems associated to the
same 4d theory, whose moduli space metrics must somehow
coincide. We'll illustrate by examples how these systems
correspond to a different choice of a surface operator in the
same 4d gauge theory.

The construction is extended in \cite{Hanany:1997vm}: one can
add a D2 branes (along the $017$ directions) attached to one
NS5 brane in order to produce interesting 2d sigma models. As
remarked in \cite{Alday:2009fs}, the construction is actually
producing a surface operator. If the 4d theory engineered by
the brane setup is trivial, the degrees of freedom living on
the surface operator describe a purely 2d theory.

\subsection{The $\IC \IP^1$ sigma model}
The $\IC \IP^1$ sigma model is the canonical example of a
massive $(2,2)$ theory in 2d. It can be described neatly by a
linear sigma model, with a $U(1)$ gauge field in two dimensions
coupled to two chiral fields $q^i$ of charge $+1$. The only
protected coupling is the complexified FI parameters $t$ for
the $U(1)$ gauge field. It determines the size of the $\IC
\IP^1$ target space by the SUSY constraints $\sum_i |q^i|^2 =
t$. It is renormalized at one loop, so that $\exp 2 \pi i t$
has the dimension of a strong coupling scale squared.

The $U(1)$ gauge symmetry is Higgsed, and eats the overall
phase of the $q^i$, leading to the $\IC \IP^1$ sigma model.
Notice that the mass of the gauge boson is of order $g_{YM}
|t|$, so the linear sigma model is an arbitrarily good
description of the $\IC \IP^1$ sigma model as $g_{YM}$ is made
very large. $g_{YM}$ is not a protected coupling, and does not
affect the protected quantities we are interested in.

There is also a $SU(2)$ flavor symmetry acting on the two
chiral fields. In general, we can turn on a twisted mass
parameter $m$ in the Cartan of the $SU(2)$. This can be defined as an expectation
value for the scalar component of a background vector multiplet, gauging the $SU(2)$ flavor symmetry.
If the mass parameter is sufficiently large, the theory is weakly coupled:
the massive chiral fields can be integrated out, and one is
left with an effective twisted superpotential for the $U(1)$
gauge field scalar partner $x$:
\begin{equation}
-2 \pi i t x + (x-m) \left[\log (x-m)-1\right] + (x+m) \left[\log (x+m)-1\right]
\end{equation}

The twisted chiral ring equation is then $ 2 \pi i t  =
\log(x-m) + \log (x+m)$ or $x^2 = m^2 + e^{2 \pi i t}$. This
result is actually valid for all values of $m,t$. The parameter
space ${\cal P}$ is a cylinder parameterized by $t$. The space
$\hat {\cal P}$ is the curve defined by the equation $x^2 = m^2
+ e^{2 \pi i t}$, and the canonical differential is $\lambda =
x dt$. The $e^{2 \pi i t}$ correction to the $x^2 = m^2$ classical twisted chiral ring
can be seen as a 2d 1-instanton effect.

This is an example of a model with a six dimensional
construction. The authors of \cite{Hori:2000ck} engineered the
model with a IIA brane construction (see fig. {fig:cp1}): two
semi-infinite D4 branes ending on the same side of an NS5
brane, and a D2 brane ending on the system. The brane
configuration can be lifted to M-theory and reduced to a simple
six-dimensional engineering construction, based on the $A_1$ 6d
SCFT. \cite{Gaiotto:2009hg}, \cite{Gaiotto:2009we}. The theory
is compactified on a cylinder (or, equivalently, the two
punctured sphere), with boundary conditions encoded by the
quadratic differential
\begin{equation}\phi_2 = (m^2 + e^{2 \pi i t}) dt^2 =(\frac{m^2}{z^2} + \frac{\Lambda^2}{z}) dz^2 \end{equation}
The second expression is suitable for the two punctured sphere,
we defined $e^{2 \pi i t} = \Lambda^2 z$ in terms of a scale
$\Lambda$ and a dimensionless parameter $z$. The D2 brane goes
to a minimal surface operator in the setup. In a sense, this
construction gives the 2d sigma model as a surface operator in
a trivial 4d theory.

\begin{figure}
  \begin{center}
    \includegraphics[width=5in]{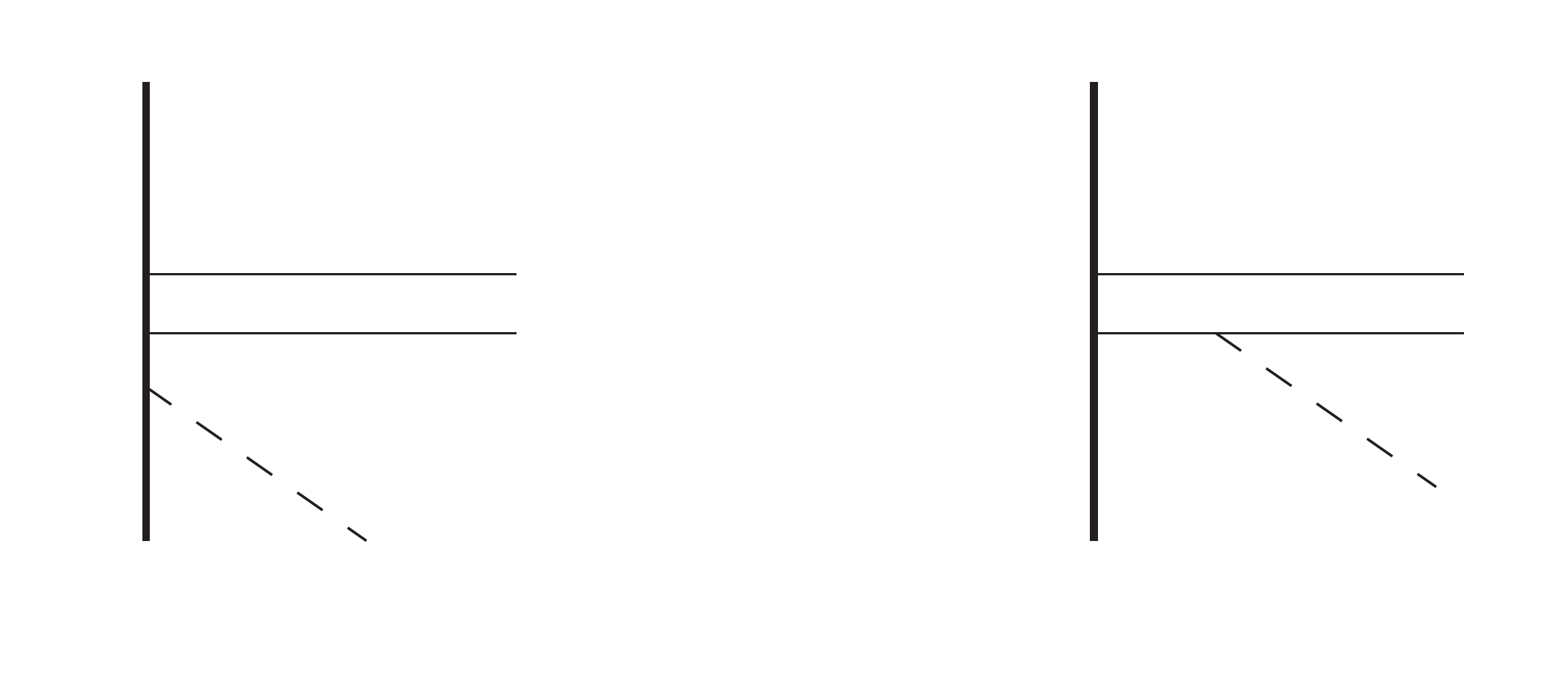}
      \end{center}
  \caption{A brane realization of the $\IC \IP^1$ sigma model. The dashed line represents a D2 brane ending on the system.
  If the brane ends on the NS5 brane, the linear sigma model can be recovered. Turning on an FI term moves the D2 along the D4 branes.}
  \label{fig:cp1}
\end{figure}

The $tt^*$ equations for the model correspond to a $SU(2)$
Hitchin system with a regular singularity at $z=0$ and an
irregular singularity at $z=\infty$. The quadratic differential
$\phi_2$ has no normalizable deformation, and the moduli space
of solutions to the Hitchin system is zero-dimensional.

\subsection{The minimal surface operator in $SU(2)$ Seiberg-Witten theory}
The $SU(2)$ Seiberg-Witten theory can be engineered in Type IIA
string theory by two D4 brane segments suspended between a pair
of NS5 branes. The basic six dimensional engineering involve
the compactification of the $A_1$ 6d SCFT on a cylinder, with
boundary conditions at the ends corresponding to the quadratic
differential \cite{Gaiotto:2009hg}
\begin{equation}\phi_2 = (\frac{\Lambda^2}{z^3} + \frac{2u }{z^2} + \frac{\Lambda^2}{z}) dz^2 \end{equation}
Here $u$ is the Coulomb branch parameter, expectation value of
$\mathrm{Tr} \Phi^2$, where $\Phi$ is the adjoint scalar
superpartner of the $SU(2)$ gauge field. The scale $\Lambda$ is
related to the UV renormalized gauge coupling $\Lambda^4 = \exp
2 \pi i \tau$. The coordinate $z$ is also the parameter of a
minimal surface operator. At weak 4d coupling, $|u|\geq
|\Lambda^2|$, there are are three interesting ranges of values
for $z$, depending on which of the three terms in $\phi_2$
dominates.

If $z$ is of order one, $\lambda \sim \frac{a}{z}$ with small
corrections of order $\frac{\Lambda^2}{a}$. (At weak coupling,
$2 u \sim a^2$). The IR effective superpotential has the form
$a \log z$. The IR couplings take the form of $2 \pi i t^{IR}
\sim \log z = 2 \pi i t $, roughly $a$ independent up to the
order of $\frac{\Lambda^2}{a^2}$ corrections. The surface
operator in this intermediate regime for $z$ is well described
by the definition as a gauge theory defect. The parameter $t$
lives on a cylinder, rather than the expected torus, because
$t=\eta + \tau \alpha$, but $\tau$ diverges at short distances,
where the defect is defined.

\begin{figure}
  \begin{center}
    \includegraphics[width=5in]{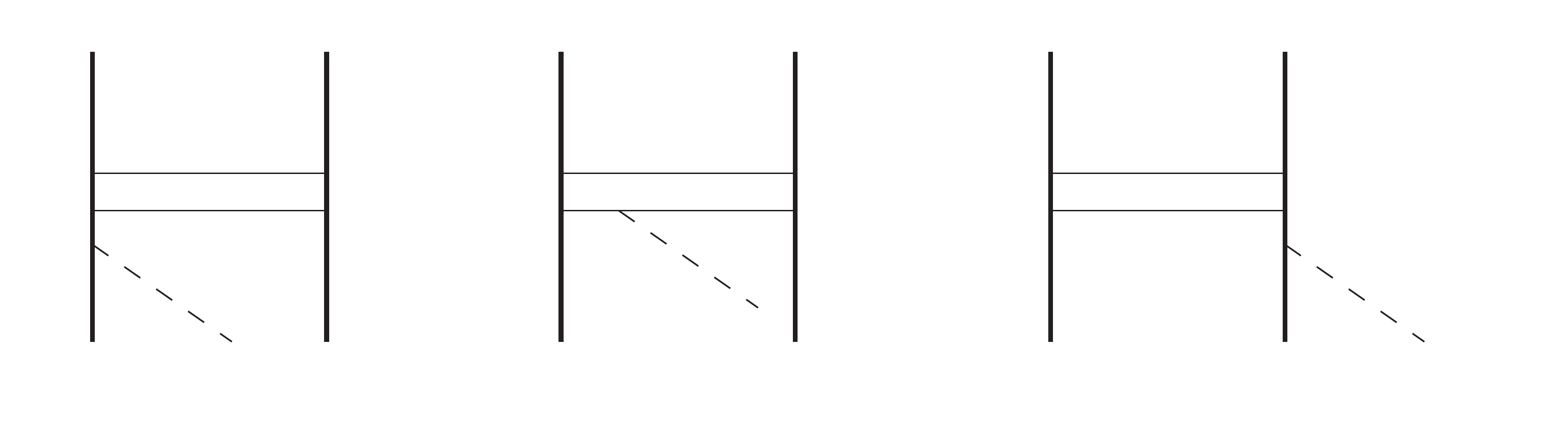}
      \end{center}
  \caption{A brane realization of the minimal surface operator in the pure $SU(2)$ theory.
  If the D2 brane ends on the NS5 brane, the description as coupling to a $\IC \IP^1$ sigma model is recovered. If the D2 ends on the D4 the realization as a defect in the gauge theory is recovered.}
  \label{fig:su2nf0}
\end{figure}

As $z$ becomes sufficiently large or sufficiently small, the
first or the last terms of the quadratic differential dominate.
In terms of the IIA brane picture, the minimal surface operator
is exploring the region near either NS5 brane, where the system
resembles the one used to engineer an $\IC\IP^1$ sigma model.
(See fig. \ref{fig:su2nf0}) For large $z$, it is instructive to
use a coordinate $\Lambda^2 z = e^{2 \pi i t}$, to get
\begin{equation}\phi_2 = (\Lambda^4 e^{-2 \pi i t} + 2u + e^{2 \pi i t}) dt^2 \end{equation}
This indeed resembles the one for an $\IC\IP^1$ sigma model,
coupled to the 4d $SU(2)$ gauge group. The extra correction
$\Lambda^4 e^{-2 \pi i t}$ seems to have a simple physical
interpretation: a combination of a 2d-4d instanton with 4d
instanton number $1$, and 2d instanton number $-1$. It would be
interesting to compute this term directly in field theory, and
understand in detail how the presence of the 4d instanton
allows for a negative 2d instanton number.

\subsection{Non-minimal surface operators in $SU(2)$ Seiberg-Witten theory: Coupling to an $\IC \IP^2$ sigma model.}

Now, consider again the brane configuration engineering pure
the $SU(2)$ theory and the following manipulation: add a D6
brane to the left of all NS5, D4 branes, and them move it to
the right, taking care to follow the appropriate rules of D4
brane creation due to the Hanani-Witten effect. Each time the
D6 crosses an NS5 a new D4 brane segment appears. As a result,
we are left with three D4 brane segments between the two NS5
branes, and two between the rightmost NS5 and the D6 brane. The
SW curve takes a form
\begin{equation}
\Lambda^3 z^2 + (x-m)(x^2-2u) z + \Lambda (x-m)^2 =0
\end{equation}
$m$ is the transverse position of the D6 brane. The $(x-m)$
factors represent the new D4 brane segments attached to the D6
brane.  The SW differential is $\lambda = x \frac{dz}{z}$. In
practice, this is the usual $SU(2)$ curve, subject to a
coordinate redefinition $z \to z \Lambda(x-m)^{-1}$, which does
not change the form of the SW differential.

\begin{figure}
  \begin{center}
    \includegraphics[width=5in]{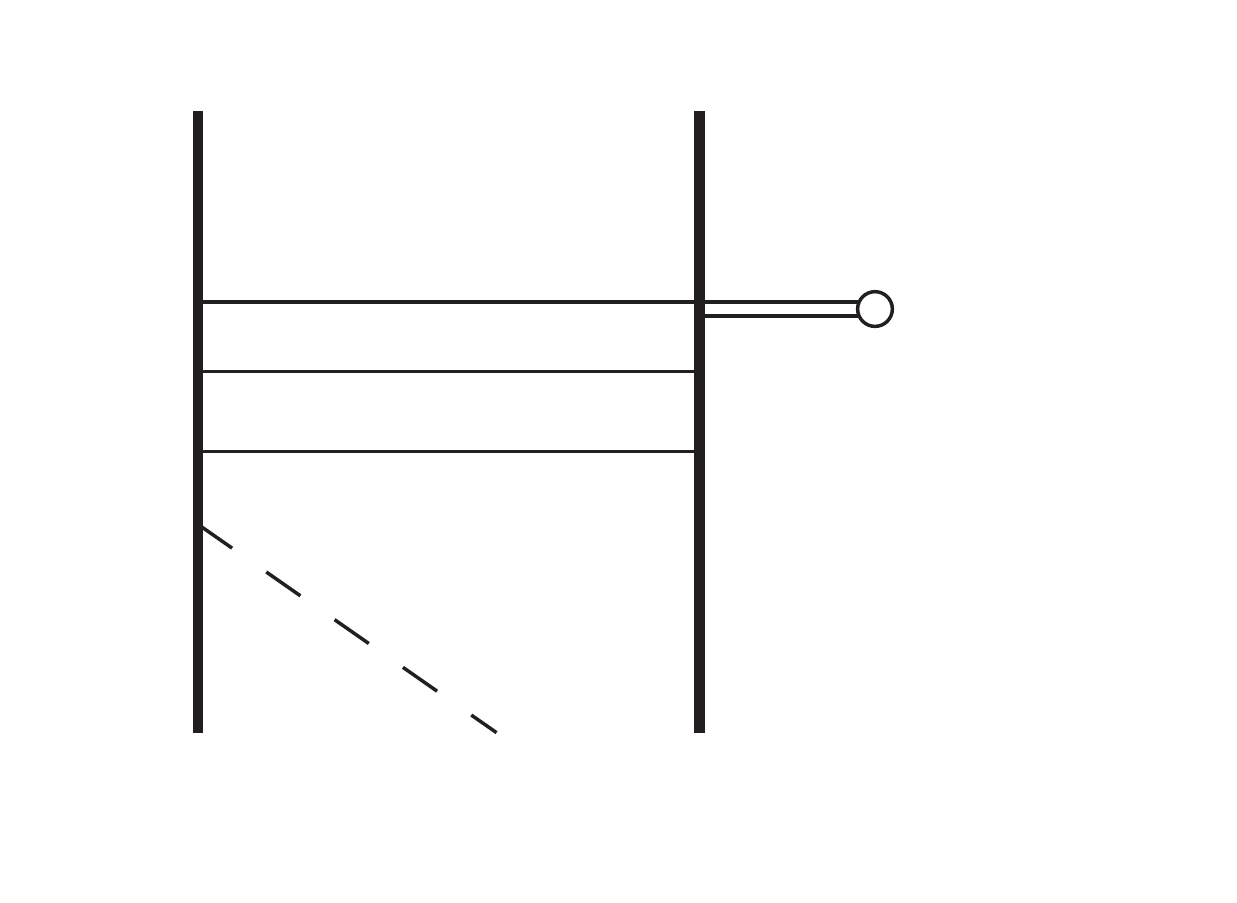}
      \end{center}
  \caption{A brane realization of a non-minimal surface operator in the pure $SU(2)$ theory.}
  \label{fig:su2nf0bis}
\end{figure}

As we have now three D4 segments ending on the leftmost NS5
brane (see fig. \ref{fig:su2nf0bis}), by \cite{Hanany:1997vm}
construction we expect that a D2 brane ending on that NS5 brane
will give rise to a $\IC \IP^2$ 2d sigma model. Indeed at weak
4d gauge coupling $2u \sim a^2 \geq \Lambda^2$ and large $z$
the curve approaches
\begin{equation}
\Lambda^3 z + (x-m)(x^2-a^2) =0
\end{equation}
which is the correct curve for a $\IC \IP^2$ 2d sigma model
with twisted mass parameters $(m,a,-a)$. We see that the 4d
$SU(2)$ gauge group is embedded in the $SU(3)$ flavor symmetry
by decomposing the fundamental $3 \to 2+1$. This embedding
commutes with a residual $U(1)$ flavor symmetry, associated to
the $m$ mass parameter.

\subsection{Non-minimal surface operators in $SU(2)$ Seiberg-Witten theory: Coupling to an $\IC \IP^n$ sigma model.}
The above example can be easily generalized to surface
operators defined by a coupling of pure $SU(2)$ gauge theory to
$\IC \IP^{n-1}$ sigma models, as long as the $SU(2)$ gauge
group is embedded as a $2 \times 2$ block in the $SU(n)$ flavor
group.

If we simply carry $n-2$ D6 branes from left to right, the SW
curve takes a form
\begin{equation}
\Lambda^n z^2 + (x^2-2u) z \prod^{n-2}_{i=1}(x-m_i) + \Lambda^{4-n} \prod^{n-2}_{i=1}(x-m_i)^2 =0
\end{equation}

This is the spectral curve for an $SU(2n-4)$ Hitchin system. It
is curious that the surface operator appears to have $n-4$
extra vacua besides the ones of the $\IC \IP^{n-1}$ sigma
model. It would be interesting to explore the dynamics of this
system.

\subsection{Non-minimal surface operators in $SU(2)$ Seiberg-Witten theory: Triplet coupling to an $\IC \IP^2$ sigma model.}
A second natural way to embed $SU(2)$ in $SU(3)$ is to embed it
as $SO(3)$, i.e. to use the triplet representation of $SU(2)$.
Hence it should be possible to consider a pure $SU(2)$ gauge
theory coupled to a $\IC \IP^2$ sigma model this way. It is not
quite obvious how to realize this in a brane setup. As we want
$SU(2)$ to act as a triplet, we are tempted to take the
Seiberg-Witten curve, written as a spectral curve for an
$SU(2)$ Hitchin system, i.e. as a determinant in the
fundamental representation $\det_{(2)} (x-\phi(z) )=0$, and
rewrite it as a determinant in the triplet representation
\begin{equation}
\det_{(3)} (x-\phi(z))= x(x^2 + 4 z - 8 u + 4 z^{-1}) = 4 z x + x(x^2- 8 u) + 4 z^{-1} x=0
\end{equation}
The Seiberg-Witten differential is the usual $x \frac{dz}{z}$.
As this resembles the curve derived from the set of branes in
figure \ref{}, albeit with restrictions on the allowed values
of the parameters, let us assume we are allowed to mimick the
D6 brane manipulations in the figure, to arrive to a simple
proposal
\begin{equation}
4 z + x(x^2- 8 u) + 4 z^{-1} x^2 = x^3 + 4 z^{-1} x^2 - 8 u x + 4 z=0
\end{equation}

The large $z$ behavior of this curve reproduces the expected
behavior of the $\IC \IP^2$ sigma model. This is the spectral
curve for a rather reasonable $U(3)$ Hitchin system. The
constraint on the coefficient is actually rather simple: the
curve is symmetric under $x \to -x$ and $z \to -z$. At this
point, we recognize an actual six-dimensional configuration:
the $A_2$ six dimensional SCFT has a $Z_2$ outer automorphism
under which the protected operators of odd degree are odd. This
is a compactification of the $A_2$ theory on a cylinder of
coordinate $\tilde z = z^2$, with a twist line for the outer
automorphism along the cylinder and appropriate boundary
conditions at the two ends. Going from M-theory to Type IIA by
reduction on the circle in the cylinder, it is known that such
a twist line will give rise to an $SO(3)$ gauge theory. This
brane construction could be a way to justify our proposed
curve.

\subsection{Surface operators and flavors: $N_f=1$ $SU(2)$ Seiberg-Witten theory}
The $N_f=1$ $SU(2)$ Seiberg-Witten theory can be engineered in
Type IIA string theory by two D4 brane segments suspended
between a pair of NS5 branes, together with an extra
semi-infinite D4 brane at either end. The six dimensional
engineering involve the compactification of the $A_1$ 6d SCFT
on a cylinder, with boundary conditions at the ends
corresponding to the quadratic differential
\begin{equation}\phi_2 = (\frac{\Lambda^2}{z^4} +\frac{2 m \Lambda}{z^3} + \frac{2u }{z^2} + \frac{\Lambda^2}{z}) dz^2 \end{equation}
Here $u$ is the Coulomb branch parameter, the scale $\Lambda$
is related to the UV renormalized gauge coupling $\Lambda^3 =
\exp 2 \pi i \tau$, $m$ is the mass parameter for the single
flavor of the theory. The coordinate $z$ is also the parameter
of a minimal surface operator. Again, there are are various
interesting ranges of values for $z$.

The most notable point is the asymmetry between the two ends of
the cylinder parameterized by $z$. At weak coupling, and
intermediate values of $z$, we can again describe the surface
operator rather well as a defect. The endpoints of the cylinder
corresponds to the values of the monodromy parameter $\alpha$
at which the monodromy vanishes in the adjoint representation.
In the presence of matter in a fundamental representation, the
two ends will correspond either to a trivial monodromy or to
antiperiodic boundary conditions for the matter hypermultiplet.

When the surface operator moves to either end of the cylinder,
in the type IIA description we can describe it as a D2 brane
ending on either NS5 branes. At large $z$ the D2 brane only
communicates with the two finite D4 brane segments, and the
theory at the defect is a $U(1)$ linear sigma model with two
chiral multiplets of charge $1$, i.e. a $\IC \IP^1$ sigma
model. Indeed at large $\Lambda^2 z = e^{2 \pi i t}$ we see the
$\IC \IP^1$ chiral ring relation with interesting 4d instanton
corrections.
\begin{equation}x^2 = e^{2 \pi i t} + 2 u + 2 m \Lambda^3 e^{-2 \pi i t} + \Lambda^6 e^{-4 \pi i t}\end{equation}

On the other hand, at small $z$, the D2 brane interacts with
the semi-infinite D4 brane as well, giving rise to a $U(1)$
linear sigma model with two chiral multiplets $q^i$ of charge
$-1$ and a single one $\tilde q$ of charge $1$. This is a sigma
model with a non-compact target space $O(-1) \to \IC \IP^1$.
One might be troubled by the non-compactness of the target
space, but there is an allowed superpotential coupling between
the matter hypermultiplet and the 2d sigma model, of the form
\begin{equation}
\tilde q Q_i q^i
\end{equation}
This coupling is marginal, but probably not exactly marginal.
This superpotential forces the identification between the
flavor symmetry of the 4d hypermultiplet and the flavor
symmetry acting on $\tilde q$ (and of the corresponding mass
parameters). Any expectation value for $\tilde q$ would source
the 4d hypermultiplets. It would be interesting to understand
in better detail the effect of this term.

%In any case, the chiral ring relation for the linear sigma
%model takes the form $2(x+m) e^{2 \pi i t} = x^2 - a^2$, which
%after a shift of $x$ by $e^{2 \pi i t}$ gives
%\begin{equation}x^2 = e^{4 \pi i t} + 2 m e^{2 \pi i t} + a^2\end{equation}
%After $z \to \Lambda e^{-2 \pi i t}$ we recognize the small $z$
%behavior of the quadratic differential.

\subsection{Surface operators and flavors: $N_f=2$ $SU(2)$ Seiberg-Witten theory}
The $N_f=2$ $SU(2)$ Seiberg-Witten theory can be engineered in
Type IIA string theory by two D4 brane segments suspended
between a pair of NS5 branes, together with two extra
semi-infinite D4 brane at either end. There are two distinct
choices: two semi-infinite D4 branes at the same end, or at
different ends.

The first choice leads to the $A_1$ 6d SCFT compactified on a
cylinder, with boundary conditions at both ends resembling the
$N_f=1$ case:
\begin{equation}\phi_2 = \left(\frac{\Lambda^2}{z^4} +\frac{2 m \Lambda}{z^3} + \frac{2u }{z^2} + \frac{2\Lambda m'}{z} + \Lambda^2\right) dz^2 \end{equation}
Here $u$ is the Coulomb branch parameter, the scale $\Lambda$
is related to the UV renormalized gauge coupling $\Lambda^2 =
\exp 2 \pi i \tau$, $m$, $m'$ are the mass parameters for the
two flavors of the theory. It should be rather clear that the
minimal surface operator in this setup must be treating the two
flavor hypermultiplets in an asymmetric way. In the defect
description, the simplest possibility is that the monodromy of
the two fundamental hypers differs by a factor of $-1$. This
would break the $SO(4)$ flavor symmetry to $SO(2) \times
SO(2)$, which is the flavor symmetry manifest in the
six-dimensional construction. At both ends, we see a
description in terms of the $O(-1) \to \IC \IP^1$ sigma model.

The second choice leads to a compactification with three
punctures, two regular ones and an irregular one.
\begin{equation}\phi_2 = \left(\frac{m_1^2}{z^2} + \frac{m_2^2}{(z-1)^2} + \frac{2 u}{z(z-1)} + \frac{\Lambda^2}{z}\right) dz^2 \end{equation}
Each regular puncture carries an $SU(2)$ flavor group,
realizing the full $SO(4)$ flavor group of the theory. This
indicates that the minimal surface operator in this
construction treats the two 4d hypermultiplets symmetrically,
and that the defect description involves the same monodromy
parameter for the two. Near the irregular puncture the behavior
of the quadratic differential is the same as near the ends of
the $N_f=0$ cylinder. The brane construction confirms the
description as a simple $\IC \IP^1$ sigma model when the D2
brane is near the NS5 brane without semi-infinite D4 branes.

On the other end of the brane system, as observed in
\cite{Alday:2009fs}, we get a $O(-2) \to \IC \IP^1$ sigma
model, i.e. a conifold sigma model. As discussed in
\cite{Alday:2009fs}, this description of the surface operator
worldvolume theory is a bit confusing, as the superpotential
coupling only preserves an $U(2)$ subgroup of the $SO(4)$
flavor symmetry. In a sense, the 2d sigma model is a
description which is valid near one of the regular punctures,
where the surface operator takes the form of a defect for one
of the flavor symmetry groups.

\subsection{Surface operators and flavors: $N_f=3$ $SU(2)$ Seiberg-Witten theory}
If we consider a surface operator which treats the three
flavors in a symmetric fashion, we encounter an interesting
phenomenon. The IIA brane setup leads to a six dimensional
construction involving three M5 branes. In the intermediate
region of parameters where the defect description should be
appropriate, two of the M5 branes represent the D4 brane
segments over which the $SU(2)$ gauge theory lives. The third
M5 brane represents one of the NS5 branes, which is bent
inwards by the pull of the three semi-infinite D4 branes. In
particular, the defect theory must three vacua! Two are the
usual perturbative ones, but the third vacuum must be
non-perturbative. In the third vacuum, the $SU(2)$ gauge
symmetry is somehow restored at the defect. It would be
interesting to explore the physical meaning of this fact.

\subsection{Surface operators and product theories: $SU(2) \times SU(2)$ Seiberg-Witten theory}
Finally, we would like to find an example of a surface operator
coupling two otherwise decoupled bulk theories. We aim to
describe the coupling of an $\IC \IP^3$ sigma model to a pair
of pure $SU(2)$ gauge theories in the bulk, embedded in a
block-diagonal fashion inside the $SU(4)$ flavor symmetry of
the sigma model. The 4d gauge group commutes with a diagonal
$U(1)$ flavor symmetry. We will borrow a construction from
\cite{Kapustin:1998fa}, where orbifold planes were added to the
standard D4-NS5 construction in order to engineer quivers of
unitary groups in the shape of D-type Dynkin diagrams. The
M-theori lift of the orbifold plane leads to M5 branes wrapping
a orbifolded Riemann surface C, either a cylinder or a torus.
There is a $Z_2$ action $t \to -t$ (or $z \to 1/z$), with two
fixed points.

In order to engineer a pair of $SU(2)$ groups, we want to use an $SO(4)$ Dynkin diagram,
which is reproduced by a single orbifold plane, in the presence of two NS5 branes (and their mirror images),
and four D4 brane segments stretched between the furthest NS5 brane and its mirror image.
The orbifold plane restricts the D4 branes to break according to a certain pattern (see fig. 6 in the reference)
so that one really has two sets of two D4 brane segments, leading to two decoupled  $SU(2)$ theories.
The counting of degrees of freedom is quite manifest in the SW curve.
One starts with a
\begin{equation}
z^2 + P_4(x) z + Q_4(x) + P_4(x)/z + 1/z^2 =0
\end{equation}
but has to impose the extra constraint on the degree $4$ polynomials $P_4$, $Q_4$ that
at the fixed points $z=\pm1$ the roots of $x$ should all be double.
Hence we can write $P_4(x) = p_2(x) q_2(x)$, $Q_4(x) = p_2(x)^2 + q_2(x)^2-2$,
so that at $z=\pm1$ the equation becomes $(p_2(x) \pm q_2(x))^2=0$.
Without loss of generality, we can set $\Lambda_1 p_2(x) = (x+m)^2-2u_1$,
$\Lambda_2 p_2(x) = (x-m)^2-2u_2$.

We claim that $\Lambda_i$, $u_i$ are the parameters of the two
$SU(2)$ theories, and $m$ is the 2d flavor symmetry mass
parameter. We do not know of a simple proof that the periods of
this SW curve coincide with linear combinations of the periods
of the two $SU(2)$ theories and $m$. We checked this result at
the first few orders of a weak coupling power expansion. It
would be nice to execute a full check through the appropriate
Picard-Lefschetz equations. At large $z$, we see the curve for
the $\IC \IP^3$ sigma model.

\section{Final remarks} \label{sec:conclusions}
Consider the following setup: pick any choice of 4d theory and any one-parameter surface operator with, say,
an  $SU(2)$ flavor symmetry. If we couple that $SU(2)$ flavor symmetry to an extra $SU(2)$
Seiberg-Witten theory, something remarkable will happen. The curves ${\cal P}$, $\hat {\cal P}$
will be modified in a rather interesting way,
to accomodate the periods of the pure $SU(2)$ theory, and the new Hitchin system
must have a moduli space which is the fibered product of the old moduli space and
the moduli space of the pure $SU(2)$ theory (the product is fibered because the $u$ parameter
of the pure $SU(2)$ theory determines the mass parameter of the old theory).

It is quite astonishing that such an universal transformation should always be possible.
This hints to the existence of some interesting underlying mathematical structure.
One could explore, for example, if the Picard-Lefschetz equations satisfied by the
periods are modified in some standard fashion.

In this paper we have not addressed the duality between the instanton partition function of
4d theories and 2d conformal blocks. While we expect the duality to also admit a formulation
in terms of surface operators, we do not know how could it be extended to situations where
the surface operator parameter space has dimension higher than one. For that to be possible,
one would need to formulate some sort of higher dimensional generalization of, say,
Liouville theory. It would have to be partly topological, as it should only depend on the
complex structure of ${\cal P}$.

\section*{Acknowledgments}
The author has benefited from discussions with S.Cecotti,
G.Moore, A.Neitzke,Y.Tachikawa, C.Vafa, H. Verlinde, E.Witten.
D.G. is supported in part by the NSF grant PHY-0503584. D.G. is
supported in part by the Roger Dashen membership in the
Institute for Advanced Study.

\appendix
\section{A review of $tt^*$} \label{app:tt}
In this appendix we will review the derivation of the $tt^*$
equations of \cite{Cecotti:1991me}, and combine it in a useful
way with some important results derived in \cite{Hori:2000ck}.
A slight difference from the standard derivation is that we
will make less the use of topological twists. We plan to use
$tt^*$ equations for surface operators in ${\cal N}=2$ four
dimensional theories. It is unclear if a useful topological
twist of the combined 4d-2d system exists which reduces to the
standard twist of $(2,2)$ theories in 2d. Donaldson-Witten
twist probably does the job, but as the surface operator breaks
$SU(2)_R$ of the bulk theory to an $U(1)$ subgroup, the 4d
manifold has to be Kahler. Instead of checking at each step if
the properties of the 2d topological twist can hold for surface
operators, we prefer to work directly with the physical theory.
In particular, we only consider the physical parameter space of
the theory, rather than the topological extended parameter
space: we only consider deformations by exactly marginal
operators. We also minimize the use of the ``topological
basis'' of Ramond vacua.

The $tt^*$ equations govern the behavior of the Ramond vacua of
a massive 2d theory with $(2,2)$ SUSY compactified on a circle.
As in the main text, we will denote the left moving
supercharges as $Q_L$ and $\bar Q_L$, the right moving
supercharges as $Q_R$ and $\bar Q_R$. $Q_L$ and $Q_R$ have
opposite $U(1)_V$ charge.

We assume the theory has an exactly marginal parameter space
${\cal P}$. The theory defined on a line has a finite set of
Lorentz invariant vacua, which can be fibered over ${\cal P}$
to give a ramified cover $\hat {\cal P}$. The vacua are
parameterized by the expectation values of operators in the
twisted chiral ring. A particularly interesting set of twisted
chiral operators $\hat x_a$ generates deformations of the
parameters $z^a \to z^a + \delta z^a$ along ${\cal P}$, by
adding a the twisted 2d superpotential $x_a \delta z^a$ to the
theory. The deformation parameter $\delta z^a$ can be taken to
be infinitesimal or finite. In the latter case the expression
$z^a + \delta z^a$ should be intended as the change in the
parameters due to the finite change in the twisted
superpotential. The $x_a$ may or may not generate the whole
twisted chiral ring, so their expectation values may or may not
be sufficient to separate the vacua of the theory. In any case,
they give a projection of $\hat {\cal P}$ to a ramified cover
of ${\cal P}$ inside $T^*{\cal P}$.

The Ramond vacua of the theory compactified on a circle
$S^1_\beta$ define a hermitean vector bundle ${\cal V}$ on
${\cal P}$. Ramond vacua are killed by all supercharges. The
$tt^*$ equations involve certain natural connections on ${\cal
V}$. In order to define a connection on  ${\cal V}$, we need to
be able to compare vacua of the theory at close but distinct
values of the parameters. An example of such a comparison is a
path integral on a cylinder geometry, with asymptotic value of
the parameters $z^a$ on one side, $z^a+ \delta z^a$ on the
other side, and at the two ends appropriate choices of vacua
$\langle v,z|$ and $| v', z^a+ \delta z^a\rangle$. If we can
make sense of the change of $z$ between the two ends, this
defines an inner product
\begin{equation}
\langle v,z| v', z^a+ \delta z^a\rangle
\end{equation}
and hence a connection.

To this end we need to understand how to define the theory on a
line, with different asymptotic values of the parameters $z^a$,
$z^a+ \delta z^a$. This is a 2d example of a ``Janus domain
wall''. A naive possibility is to simply add a position
dependent twisted superpotential term
\begin{equation}
\int d\sigma d \tau f(\sigma) \{Q_R,\{Q_L,x\}\}+\bar f(\sigma) \{\bar Q_R,\{\bar Q_L,\bar x\}\}
\end{equation}
This term does the job and is hermitean, but breaks all
supersymmetries. At the first order of perturbation theory in
$f,\bar f$ the cylinder calculation in this background is
sensible, and gives a hermitean connection $D_a, \bar D_a$.
Indeed, for a change of the profile $\delta f$ which goes to
zero sufficiently fast the change in the twisted superpotential
\begin{equation} \int
d\sigma d \tau \delta f(\sigma) \{Q_R,\{Q_L,x\}\}+\bar \delta f(\sigma)
\{\bar Q_R,\{\bar Q_L,\bar x\}\}
\end{equation}
is a sum of Q-exact terms, which annihilate the Ramond vacua at
either ends of the cylinder, without any chance of interesting
contact terms. Hence the answer only depends on the value of
the parameters at the two ends of the cylinder.

On the other hand, for finite $f, \bar f$, the result depends
on the specific profile: if we vary the profile, we add Q-exact
terms in s SUSY breaking background. In particular, the
hermitean connection $D_a, \bar D_a$ has no reason to be flat.
If we turn off the anti-holomorphis $\bar f$ deformation
parameter though, the superpotential term preserves both $Q_R$
and $Q_L$, and the variations in the profile $\delta f$ are
$Q_R$ and $Q_L$ exact, hence the ``holomorphic'' Janus domain
wall gives a well defined transport in holomorphic directions:
the $(2,0)$ part of the connection $[D_a,D_b]$ vanishes. The
same is true for $[\bar D_a, \bar D_b]$

To compute the $(1,1)$ part of the connection, we need to be
able to deal with both holomorphic and antiholomorphic
deformations. To archive that, we will now define a
supersymmetric Janus wall, preserving the interesting
combinations $\zeta Q_L + \bar Q_R$, $\zeta Q_R + \bar Q_L$.
\begin{equation}
\int d\sigma d \tau \left[ f(\sigma) \{Q_R,\{Q_L,x\}\}+ \zeta^{-1} \partial_\sigma f x \right] +\left[\bar f(\sigma) \{\bar Q_R,\{\bar Q_L,\bar x\}\} + \zeta \partial_\sigma \bar f \bar x \right]
\end{equation}
Indeed if we act with $Q_L + \zeta \bar Q_R$ we get
\begin{equation}
\int d\sigma d \tau \left[ f(\sigma) \partial_R \{Q_L,x\}+ \partial_\sigma f \{Q_L,x\} \right] +\left[\bar f(\sigma) \zeta \partial_L \{\bar Q_R,\bar x\} + \zeta \partial_\sigma \bar f \{\bar Q_R,\bar x\} \right]
\end{equation}
which integrates by parts to zero. Furthermore, variations of
the profile are $\zeta Q_L + \bar Q_R$ or $\zeta Q_R + \bar
Q_L$ exact.

This supersymmetric Janus wrapped on a cylinder defines a flat
connection $\nabla_a, \bar \nabla_a$ which depends on the
spectral parameter $\zeta$. This connection can be written in
terms of $D_a, \bar D_a$ and of the matrices $c_a$ which give
the action of $x_a$ on the Ramond vacua of the theory
\begin{equation}
\nabla_a = D_a + \frac{\beta}{\zeta} c_a \qquad \bar \nabla_a = \bar D_a + \beta \zeta \bar c_a
\end{equation}
Flatness of this spectral connection implies all the $tt^*$
equations, which are a sort of multidimensional generalization
of the Hitchin system on Riemann surfaces.
\begin{align}
&[D_a,D_b]=[\bar D_a, \bar D_b]=[c_a,c_b]=[\bar c_a, \bar c_b]=0 \notag \\
&D_a c_b = D_b c_a \qquad \bar D_a \bar c_b = \bar D_b \bar c_a \qquad D_a \bar c_b = 0 \qquad  \bar D_a c_b=0 \notag \\
&[D_a, \bar D_b] + \beta^2 [c_a, \bar c_b]=0
\end{align}

We can now make contact with \cite{Hori:2000ck}, where a simple
result is proven: the ``boundary entropy'', i.e. the pairing
between a Ramond vacuum and a boundary state, for some boundary
condition preserving $\zeta Q_L + \bar Q_R$ and $\zeta Q_R +
\bar Q_L$, is a flat section for the spectral connection
$\nabla_a, \bar \nabla_a$. Indeed, the fact that the correction
terms in the supersymmetric Janus configuration are
proportional to the derivative $\partial_\sigma f$, implies
that the Janus configuration has a well defined limit as the
profile $f$ becomes a step function: the correction terms take
the form of a boundary Lagrangian, which cancels the SUSY
variation of the bulk superpotential term. In this limit, the
Janus configuration becomes a sharp domain wall wall, which is
the same as a boundary condition for a doubled-up theory, the
product of the theory at $z$ and the theory at $z + \delta z$.

The matrix element $\langle v,z| v', z^a+ \delta z^a\rangle$ is
the boundary entropy for this boundary condition, and indeed is
killed both by the left action and by the right action of
$\nabla_a, \bar \nabla_a$. More generally, given a boundary
condition for the theory at some $z$, we can define a boundary
condition at $z + \delta z$ by adding the integral of the
twisted superpotential in the bulk, and the extra $\oint
\zeta^{-1} x + \zeta \bar x$ along the boundary to fix the SUSY
transformations. This shows directly that the boundary entropy
is a flat section for the spectral connection.

%One last observation: it is interesting to consider the large
%radius $\beta \to \infty$ limit of the flat transport along
%some path $p$ in ${\cal P}$. Naive WKB approximation proceeds
%by diagonalization of the dominant terms, proportional to $c_a,
%\bar c_a$. The eigenvalues of $c_a$ are simply the expectation
%values of $x_a$ on the vacua of the theory on the line, and
%live in $\hat {\cal P}$. The eigenvectors of the $c_a$ are
%transported along paths in $\hat {\cal P}$, and accumulate a
%WKB phase $\frac{R}{\zeta} \int x_a dz^a + R\zeta \int \bar x_a
%d\bar z^a$ (there is also an interesting finite phase which
%depends on $D_a, \bar D_a$.) A careful WKB approximation will
%select for each $p$ an ensemble of paths $\gamma_i^(p)[\zeta]
%\in \hat {\cal P}$ (lift paths in ${\cal P}$ in the homotopy
%class of $p$),  which contribute to the calculation of the
%transport between two vacua.

\bibliography{All}{}
\end{document}